\documentstyle[12pt,amssymb]{article}
\DeclareFontShape{OML}{cmm}{m}{b}{%
   <-> cmmib10}{}
\DeclareMathAlphabet{\mathbf}{OML}{cmm}{m}{b}
\DeclareSymbolFont{boldletters}{OML}{cmm}{m}{b}
\DeclareMathSymbol{\bsigma}{\mathord}{boldletters}{27}

\def\case#1#2{\textstyle{#1\over#2}\displaystyle}
\def\i{{\rm i}}

\newcommand{\be}{\begin{equation}} \newcommand{\ee}{\end{equation}}
\newcommand{\bea}{\begin{eqnarray}} \newcommand{\eea}{\end{eqnarray}}

\begin{document}

\title{Exactly solvable $su(N)$ mixed spin ladders\thanks{In honour 
of R.~J. Baxter's sixtieth birthday. Presented at the
{\em Baxter Revolution in Mathematical Physics} Conference in Canberra,
February 13-19, 2000.}}

\author{
M.~T. Batchelor, J. de Gier and M. Maslen\\
~\\
Department of Mathematics, Australian National University,\\
Canberra ACT 0200, Australia}

\date{\today}
\maketitle
\begin{abstract}
It is shown that solvable mixed spin ladder models can be constructed 
from $su(N)$ permutators. Heisenberg rung interactions appear as
chemical potential terms in the Bethe Ansatz solution. Explicit
examples given are a mixed spin-$\case{1}{2}$ spin-$1$ ladder, a 
mixed spin-$\case{1}{2}$ spin-$\case{3}{2}$ ladder and a 
spin-1 ladder with biquadratic interactions.    
\end{abstract}


\section{Introduction}

It is well known that exact solutions of realistic models in 
statistical mechanics are of immense importance. Beyond
physical insights, they provide benchmarks against which 
approximate techniques may be tested, and in some cases,
a stimulus to further research through their strong predictive 
power. 
Let us recall a quote from Baxter's book \cite{Baxter}:

\begin{quote}

Basically, I suppose the justification for studying these
lattice models is very simple: they are relevant and they
can be solved, so why not do so and see what they tell us?

\end{quote}

This is precisely the spirit of our recent work on ladder models,
which are systems of coupled quantum spin chains.
A number of exactly solved spin ladders have been 
found.\footnote{References [2-12] provide only a partial list.}
Here by exactly solved we mean integrable in the Yang-Baxter
sense, with a corresponding Bethe Ansatz 
solution.\footnote{Another class of ladder model can be constructed 
with matrix product groundstates, see for example, \cite{KM,HMT} and
references therein.}
A particularly neat construction is that given in \cite{W,BMa,BGLM}. 
There it is shown that integrable
spin models constructed from the fundamental representation of the
algebras $su(N)$, $so(N)$ and $sp(N)$, where $N=2^n$, can be
reinterpreted as $n$-leg spin-$\case{1}{2}$ ladder models.
Here we show that mixed spin integrable ladder models can be
constructed from the $su(N)$ family for any $N$. 

The 2-leg spin-$\case{1}{2}$ ladder model of Wang \cite{W} is 
of considerable interest.
It differs from the experimentally significant \cite{DR,D} 
spin-$\case{1}{2}$ Heisenberg ladder
through a four-body spin interaction, which is necessary to make
the model solvable.
Such a four-spin
interaction term has been introduced on physical grounds \cite{NT}.
In Wang's model, the
effect of this term is to shift the critical value
of the rung coupling $J$ at which the model becomes massive.
In the integrable model the Heisenberg rung coupling breaks the
underlying $su(4)$ symmetry and appears as a chemical potential
term in the Bethe Ansatz solution. 
Wang's model was shown to be part of an $su(N)$ family of ladder
models \cite{BMa}.
The phase diagram has been calculated for the 3- and 4-leg ladder
models, which include 3- and 4-leg spin tubes \cite{GBM,GB,MGB1}.
These calculations reveal magnetisation plateaus \cite{mag} in the
presence of a magnetic field \cite{GB,MGB1}.
Moreover, the exact magnetic phase diagrams are seen to be in
qualitative agreement with those of the $n$-leg Heisenberg 
ladders \cite{CHP}. 

This paper is arranged as follows. In section 2 we review the
basic ingredients of the $su(N)$ lattice models and their
Bethe Ansatz solution. Then in section 3 we construct the
related mixed spin ladder models. In section 4 we consider the
rung interactions which preserve integrability.
Some explicit examples are given in section 5.   

\section{$su(N)$ models}

We recall that an integrable spin-$S$ chain can be constructed from
a solution of the Yang-Baxter equation. Here we briefly review this
construction for the case of the $su(N)$ algebras. The Chevalley
generators in the fundamental representation of the $su(N)$ algebra
are given by  
\be
X_{\alpha}^+ = E_{\alpha,\alpha+1}, \quad X_{\alpha}^- =
E_{\alpha+1,\alpha},\quad H_{\alpha} = E_{\alpha\alpha} -
E_{\alpha+1,\alpha+1}, \label{eq:Chev}
\ee 
for $1\leq \alpha \leq N-1$. The $N\times N$ matrices
$E_{\alpha\beta}$ have a 1 in the $\alpha$th row and $\beta$th column
and zeros everywhere else. These generators satisfy the defining
relations of $su(N)$,
\be
[ X_{\alpha}^+, X_{\beta}^- ] =\delta_{\alpha\beta} H_{\alpha},\quad
[ H_{\alpha},X_{\beta}^{\pm} ] = \pm a_{\alpha\beta}X_{\beta}^{\pm},\quad
[ H_{\alpha},H_{\beta} ] = 0.
\ee
Here, $a_{\alpha\beta}$ are the Cartan matrix elements corresponding to the
$A_{N-1}$ Dynkin diagram, given by
\bea
a_{\alpha\beta} = \left\{
\arraycolsep=5mm
\begin{array}{@{}rl}
2  & \alpha=\beta\\
-1 & \alpha=\beta\pm 1\\
0 & {\rm otherwise.}
\end{array}\right. 
\eea
{}From the Chevalley generators one may construct a spin-$\case{N-1}{2}$
operator given by
\be
(S^{\pm})^{(N)} = \sum_{\alpha=1}^{N-1} \sqrt{\alpha(N-\alpha)}
X_{\alpha}^{\pm},\quad
(S^z)^{(N)} = \frac{1}{2} \sum_{\alpha=1}^{N-1} \alpha(N-\alpha) H_{\alpha},
\label{eq:spindef}
\ee
where $S^{\pm} = S^x \pm \i S^y$. These satisfy the $su(2)$ relations.

In terms of the $su(N)$ elements a solution of the Yang-Baxter
equation is given by  
\be
P^{(N)} = \sum_{\alpha,\beta=1}^N E_{\alpha \beta} \otimes E_{\beta
\alpha}. \label{eq:perm}
\ee
It follows that the following Hamiltonian is integrable,
\be
H = \sum_{i=1}^L P^{(N)}_{i,i+1}, \label{eq:Ham}
\ee
where $P^{(N)}_{i,j}$ acts as the permutator (\ref{eq:perm}) on the
$i$th and $j$th factor in the Hilbert space $\otimes_{i=1}^L  {\mathbb
C}^N_i$ and as the identity everywhere else. $H$ can be diagonalized
using the Bethe Ansatz. The Bethe Ansatz equations are well known
\cite{S}, and given by 
\begin{eqnarray}
\left( \frac{\lambda_j^{(1)} - \frac{\i}{2}}{\lambda_j^{(1)} +
\frac{\i}{2}} \right)^L &=& \prod_{k \neq j}^{M_1}
\frac{\lambda_j^{(1)} - \lambda_k^{(1)} - \i}{\lambda_j^{(1)} -
\lambda_k^{(1)} + \i} \prod_{k=1}^{M_2}
\frac{\lambda_j^{(1)} - \lambda_k^{(2)} + \frac{\i}{2}}{\lambda_j^{(1)} -
\lambda_k^{(2)} - \frac{\i}{2}}, \nonumber\\
\\\label{eq:BAE}
\prod_{k \neq j}^{M_r}
\frac{\lambda_j^{(r)} - \lambda_k^{(r)} - \i}{\lambda_j^{(r)} -
\lambda_k^{(r)} + \i} &=& \prod_{k=1}^{M_{r-1}}
\frac{\lambda_j^{(r)} - \lambda_k^{(r-1)} - \frac{\i}{2}}{\lambda_j^{(r)} -
\lambda_k^{(r-1)} + \frac{\i}{2}} \prod_{k=1}^{M_{r+1}}
\frac{\lambda_j^{(r)} - \lambda_k^{(r+1)} - \frac{\i}{2}}{\lambda_j^{(r)} -
\lambda_k^{(r+1)} + \frac{\i}{2}}.\nonumber     
\end{eqnarray}
Here $j = 1, \ldots, M_r$ with $r = 2, \ldots, N-1$ and $M_N=0$.
The eigenenergies of $H$ are given by
\begin{equation}
E^{\rm leg} = -\sum_{j=1}^{M_1} \frac{1}{(\lambda_j^{(1)})^2 + \frac{1}{4}}.
\label{eq:eig}
\end{equation} 

The Hamiltonian (\ref{eq:Ham}) can be interpreted as that of a
spin-$S$ chain by the identification 
\be
P^{(N)}_{i,i+1} = \sum_{\alpha=0}^{N-1} (-)^{N-1-\alpha}
\prod_{\beta \neq \alpha}^{N-1} \frac{{\bf S}^{(N)}_i \cdot {\bf
S}^{(N)}_{i+1} - x_{\beta}}{x_{\alpha} - x_{\beta}}, \label{eq:cor}
\ee
where $x_{\alpha} = \case{1}{2}\alpha(\alpha+1) -S(S+1)$ and 
$N=2S+1$ \cite{BMa}.
The components of the spin operator ${\bf S}^{(N)}$ are
defined by (\ref{eq:spindef}). In the simplest case, $S=\case{1}{2}$,
one recovers the Heisenberg model, 
\be
P^{(2)}_{i,i+1} = \case{1}{2} (1+\bsigma \cdot \bsigma),
\ee
in terms of the Pauli matrices $\bsigma$.

As an historical aside, we note that the $su(3)$ case of the Bethe
equations (7) appeared 30 years ago in a paper by Baxter, with regard 
to the Bethe Ansatz solution of a colouring problem on the honeycomb 
lattice \cite{B2}.
The $su(3)$ chain, in terms of spin-1 operators, was first
solved by Uimin \cite{U}. 

\section{Ladders}

A key point in the construction is that
for every factor $p$ of $N$, the matrix $E_{\alpha\beta}$ can be
interpreted as acting on ${\mathbb C}^p \otimes {\mathbb C}^{N/p}$,
i.e., 
\be
E^{(N)}_{\alpha\beta} = E^{(p)}_{\alpha'\beta'} \otimes
E^{(N/p)}_{\alpha''\beta''},
\ee
where $\alpha = \frac{N}{p}(\alpha'-1) + \alpha''$. It follows that
the permutator (\ref{eq:perm}) can be rewritten as
\bea
P^{(N)} &=& \sum_{\alpha',\beta'=1}^{p} \sum_{\alpha'',\beta''=1}^{N/p}
E_{\alpha'\beta'} \otimes E_{\alpha''\beta''} \otimes E_{\beta'\alpha'}
\otimes E_{\beta''\alpha''} \nonumber\\
&=& \left\{ P^{(p)} \otimes P^{(N/p)} \right\}. \label{eq:factor1}
\eea
Here, the brackets indicate that we should order the factors in the
tensor product in definition (\ref{eq:perm}) of $P^{(N)}$ according to
the first line in (\ref{eq:factor1}). Accordingly, via the correspondence
(\ref{eq:cor}), the local Hamiltonian may be interpreted as that of a
ladder with spin-$\case{p-1}{2}$ degrees of freedom on one leg and
spin-$\frac{N-p}{2p}$ on the other leg. In the case of $N=4$, $p=2$ this
amounts to
\be
H = \sum_{i=1}^L \case{1}{4} (1+\bsigma_{i,1} \cdot \bsigma_{i+1,1})
(1+\bsigma_{i,2} \cdot \bsigma_{i+1,2}).
\ee

In general, any factorization of $N$,
\be
N = \prod_{j=1}^q p_j^{m_j}, \quad \sum_{j=1}^q m_j=n, \label{eq:fact}
\ee
will give rise to an $n$-leg mixed spin ladder, with 
spin-$\frac{p_j-1}{2}$ on $m_j$ legs, with Hamiltonian
\be
H = \sum_{i=1}^L \left\{ \bigotimes_{j=1}^q \bigotimes_{k=1}^{m_j}
P^{(p_j)}_{i,i+1} \right\}. \label{eq:HamGen}
\ee  
In fact, there are $\left(\frac{n!}{m_1! \cdots m_q!}\right)$
equivalent ladders depending on the ordering of the different spin
degrees of freedom on the legs. Again, in the simple case of $N=2^n$
and $p_1=2,m_1=n$, one finds,
\be
H = \sum_{i=1}^L \case{1}{2^n} \prod_{l=1}^n (1+\bsigma_{i,l} \cdot
\bsigma_{i+1,l}).
\ee
In the following we will no longer need to specify if some factors
$p_j$ are equal and we will therefore drop the detailed notation
(\ref{eq:fact}). We will write $N=\prod_{j=1}^n p_j$ where $p_j$'s are
allowed to be the same. 

It is worth mentioning that the above procedure can also be carried
out for fermionic ladders that are obtained from a graded permutation
operator \cite{H,FK}. In such a way one may construct mixed extended
$t-J$ and Hubbard ladder models.

The simplicity of the above construction lies in the simple
factorisation (\ref{eq:factor1}) property of the permutator. It is
possible however to construct anisotropic ladder models from
$R$-matrices related to the $q$-deformed $su(N)$ algebras \cite{GBM2}.

\section{Rung interactions}

For any two factors from (\ref{eq:HamGen}), the product of their
respective spin components commutes with (\ref{eq:HamGen}).  
Indeed, it can be readily verified using the definitions (\ref{eq:Chev}),
(\ref{eq:spindef}) and (\ref{eq:perm}), that
\be
\left[ (S^a)^{(p_k)}_{i,k} \otimes (S^a)^{(p_l)}_{i,l} +
(S^a)^{(p_k)}_{i+1,k} \otimes (S^a)^{(p_l)}_{i+1,l}, \left\{
P_{i,i+1}^{(p_k)} \otimes P_{i,i+1}^{(p_l)} 
\right\}\right] =0.   
\ee 
It thus follows that one can put XYZ type interactions on the rungs
which commute with the Hamiltonian (\ref{eq:HamGen}). This means that
the ladder Hamiltonian 
\be
H = \sum_{i=1}^L \left[ \left\{ \bigotimes_{j=1}^n P^{(p_j)}_{i,i+1} \right\}
+ \sum_{j<k}^n \sum_{a=1}^3 J_a (j,k) (S^a)^{(p_j)}_i
\otimes (S^a)^{(p_k)}_i \right] \label{eq:Ham+Rung}
\ee
is integrable. A magnetic field term may be added to this
Hamiltonian without destroying the integrability.

In general the rung couplings and magnetic field appear
as chemical potential terms in the Bethe Ansatz solutions, i.e., they
do not appear in the Bethe equations (7), only in 
the eigenvalue expression (\ref{eq:eig}).
This is typical of this class of ladder model.

\section{Examples}

The result (\ref{eq:Ham+Rung}) contains previously known examples. 
For $N=2^n$ and the choice $J_a (j,k) = 2J \delta_{k,j+1}$ 
it reduces to the $n$-leg spin-$\case{1}{2}$ model \cite{BMa} 
\be
H = \sum_{i=1}^L \left[ \case{1}{2^n} \prod_{l=1}^n (1+\bsigma_{i,l} \cdot
\bsigma_{i+1,l}) + \case{1}{2} J \sum_{l=1}^n \bsigma_{i,l} \cdot
\bsigma_{i,l+1} \right].
\ee
For $n=2$ this is the model discussed by Wang \cite{W}. 
 
Another interesting example is a mixed spin ladder, with spin-$\case{1}{2}$
on one leg and spin-$1$ on the other. The Hamiltonian is given by
\be
H = \sum_{i=1}^L \left\{ \case{1}{2} ( 1+\bsigma_{i}\cdot
\bsigma_{i+1}) \left[ ({\bf S}_{i}\cdot {\bf S}_{i+1})^2 + {\bf
S}_{i}\cdot {\bf S}_{i+1} - 1 \right] + J \bsigma_{i} \cdot {\bf S}_{i}
\right\},
\ee 
where we have taken the rung interactions to be isotropic. 
This model is based on the $su(6)$ Bethe equations.
For this model the two-site rung Hamiltonian consists of a 
doublet and a quadruplet, so the model remains critical for
large rung coupling.
On the other hand, the mixed spin-$\case{1}{2}$ spin-$\case{3}{2}$
model, with Hamiltonian
\be
H = \sum_{i=1}^L \left\{ \case{1}{2} ( 1+\bsigma_{i}\cdot
\bsigma_{i+1}) \left[ \case29 ({\bf S}_{i}\cdot {\bf S}_{i+1})^3 + 
\case{11}{18} ({\bf S}_{i}\cdot {\bf S}_{i+1})^2 -
\case98 {\bf S}_{i}\cdot {\bf S}_{i+1} - \case{67}{32} \right]
+ J \bsigma_{i} \cdot {\bf S}_{i} \right\},
\ee
exhibits a transition to a massive phase at some finite
rung coupling $J$.  

The other example we mention here is the spin-$1$ ladder, with
Hamiltonian 
\be
H = \sum_{i=1}^L \left\{ \left[({\bf S}_{i,1}\cdot {\bf S}_{i+1,1})^2 + {\bf
S}_{i,1}\cdot {\bf S}_{i+1,1} - 1 \right] \left[ ({\bf S}_{i,2}\cdot {\bf
S}_{i+1,2})^2 + {\bf S}_{i,2}\cdot {\bf S}_{i+1,2} - 1 \right] + J {\bf
S}_{i,1} \cdot {\bf S}_{i,2} \right\}.
\ee 
This model becomes massive at $J_c=4$, 
with the gap opening up linearly with $J$ \cite{MBG}. 

\vskip 5mm
This work has been supported by The Australian Research Council.


\end{document}